\input{aipcheck}
\documentclass[]{aipproc}

\layoutstyle{8x11single}

\begin{document}
\newcommand{\nonu}{\nonumber}
\newcommand{\sm}{\small}
\newcommand{\noi}{\noindent}
\newcommand{\nl}{\newline}
\newcommand{\bp}{\begin{picture}}
\newcommand{\ep}{\end{picture}}
\newcommand{\bc}{\begin{center}}
\newcommand{\ec}{\end{center}}
\newcommand{\be}{\begin{equation}}
\newcommand{\ee}{\end{equation}}
\newcommand{\beal}{\begin{align}}
\newcommand{\eeal}{\end{align}}
\newcommand{\bea}{\begin{eqnarray}}
\newcommand{\eea}{\end{eqnarray}}
\newcommand{\bnabla}{\mbox{\boldmath $\nabla$}}
\newcommand{\univec}{\textbf{a}}
\newcommand{\VectorA}{\textbf{A}}
\newcommand{\Pint}

\title{On naked singularities in the extreme double Reissner-Nordstr\"{o}m solution}
\classification{04.20.Jb, 04.70.Bw, 97.60.Lf}
\keywords {Exact solution, double Reissner-Nordstr\"{o}m, singular surfaces}

\author{I. Cabrera-Munguia\footnote{cabrera@zarm.uni-bremen.de} \, and Alfredo Mac\'ias\footnote{amac@xanum.uam.mx}}
{address={Departamento de F\'isica, Universidad Aut\'onoma Metropolitana-Iztapalapa \\
 A.P. 55-534, M\'exico D.F. 09340, M\'exico} }

\begin{abstract}
We present a quite simple analytical study on the appearance or absence of naked singularities in binary systems.
As an example we consider the double Reissner-Nordstr\"{o}m solution and fix the conditions it should satisfy in order to avoid or develop singular surfaces off the axis. The proof shows that singular surfaces appear as a consequence of the presence of negative masses in the solution.
\end{abstract}
\maketitle


\section{I. Introduction}
One of the most popular exact solutions of the Einstein-Maxwell (EM) system of equations, describing massive objects endowed with electric charge, is the Majumdar-Papapetrou (MP) solution \cite{Majumdar,Papapetrou}, due to its simplicity, and thermodynamical features. The masses and charges of this solution satisfy the relation $Q_{i}=\pm M_{i}$, without regard to the distance between the sources. For a binary system the MP solution describes, in the axisymmetric case, a neutral equilibrium between the gravitational and electric forces of two electrostatic extreme black holes. It is a special case of Weyl's solution \cite{Weyl}, whose masses and charges satisfy the condition $M_{2}Q_{1}-M_{1}Q_{2}=0$. In Newton's theory, the equilibrium condition between two massive charged bodies reads  $M_{1}M_{2}-Q_{1}Q_{2}=0$.\nl

On the other hand, in the context of binary systems composed by extreme double Reissner-Nordstr\"{o}m (DRN) black holes, there exists a complementary description, known as Bonnor's solution (BS) \cite{Bonnor}. The BS represents the electrostatic analogue of the well-known Kerr-NUT solution \cite{Demianski}. It can be obtained by means of a complex continuation of the parameters \cite{Bonnor2}.

The BS describes non-equilibrium states of a two-body system composed of extreme electrostatic black holes, which interact with each other by means of a strut (conical singularity) \cite{Israel} located in between. In this interacting scenario, the charges, which are opposite in sign, are greater than the corresponding masses, i.e., $|Q_{i}|>M_{i}$ \cite{IMR}. The strut in between provides an interaction force due to the pressure, computed from the conical deficit angle, it is given by the following expression \cite{CMR}:
\be \mathcal{F}= \frac{2M_{1}M_{2}}{R^{2}-(M_{1}+M_{2})^{2}}\left[1+ \frac{2M_{1}M_{2}}{R^{2}-(M_{1}+M_{2})^{2}}\right], \qquad R>M_{1}+M_{2}.\ee

\begin{figure}[ht]
\begin{minipage}{0.49\linewidth}
\centering
\includegraphics[width=2.3cm,height=6cm]{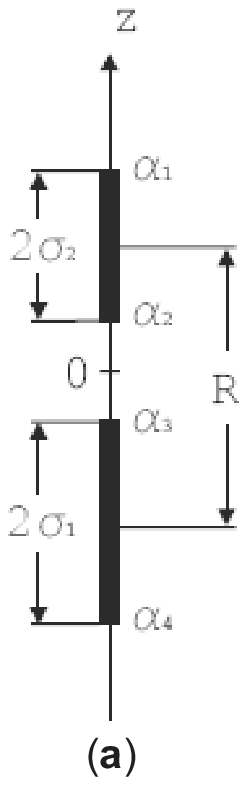}
\end{minipage}
\begin{minipage}{0.49\linewidth}
\centering
\includegraphics[width=1.4cm,height=6cm]{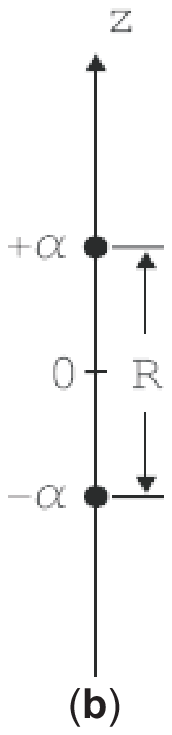}
\end{minipage}
\caption{Location of two unequal Reissner-Nordstr\"{o}m holes on the symmetry axis. (a) The non-extreme case. (b) The extreme limit case $\sigma_{j}\rightarrow0$. }
\label{DKRR}\end{figure}

It should be pointed out that the geometrical and physical properties of the extreme DRN system, can be easily described if one is able to provide a functional form, in terms of the the physical Komar parameters \cite{Komar}, of the event horizon of length $2 \sigma_{j}$ (see Fig. \ref{DKRR}). Varzugin \emph{et al.} \cite{Varzugin1} first derived the mentioned functional form for a non-extreme DRN system, in terms of the Komar masses $M_{1}$ and $M_{2}$, Komar charges $Q_{1}$ and $Q_{2}$ and a coordinate distance $R$. They read
\be \sigma_{1}=\sqrt{M_{1}^{2}-Q_{1}^{2}+2Q_{1}\frac{M_{2}Q_{1}-M_{1}Q_{2}}{R+M_{1}+M_{2}}}, \qquad
\sigma_{2}=\sqrt{M_{2}^{2}-Q_{2}^{2}-2Q_{2}\frac{M_{2}Q_{1}-M_{1}Q_{2}}{R+M_{1}+M_{2}}}, \qquad \sigma_{2}=\sigma_{1(1\leftrightarrow2)}.\label{sigmas}\ee

Eqs.(\ref{sigmas}) were used by Manko \cite{Manko} to derive the corresponding metric and physical properties of the non-extreme DRN solution in terms of the Komar parameters. Later on, the extreme DRN solution was introduced by Cabrera-Munguia \emph{et al}. \cite{IMR} as a 3-parametric exact solution resulting from a combination of the MP \cite{Majumdar,Papapetrou} and BS \cite{Bonnor} solutions in terms of canonical parameters $\{\alpha,\beta_{1},\beta_{2}\}$, where some of its physical properties were analized, in particular the relations between charges and masses arising from Eq.(\ref{sigmas}) in the extreme limit. Moreover, some numerical examples are presented for which both solutions develop naked singularities off the axis, i.e., singular surfaces (SS).

In this work we present an analytical study of the conditions under which the extreme DRN system generates or avoids SS off the axis. The well-known positive mass theorem \cite{SchoenYau1,SchoenYau2} establishes that a regular solution, i.e., a solution free of singularities, contains only allowed {\em positive} values for the total ADM mass of the system \cite{ADM}. Nevertheless, the theorem does not imply that the condition of a positive ADM total mass is enough to guarantee the regularity of the solution. Therefore, an analytical analysis of the conditions for the individual masses is needed, in order to determine if the solution is indeed non-singular, i.e., free of SS off the axis.\nl

The outline of the paper is as follows. In Sec. II the extreme DRN solution, written in terms of determinants and as a function of the canonical parameters, is presented. In Sec. III by means of a proper election of constant parameters, the MP is obtained. In Sec. IV the extreme DRN solution is reduced to the BS. We present the first analytical proof of the conditions that the solution should satisfy in order to avoid or develop SS off the axis. This proof shows that SS appear as a consequence of the presence of negative masses in the solution. In Sec. V the conclusions are presented.

\section{II. The extreme double Reissner-Nordstr\"{o}m solutions}
The extreme double Reissner-Nordstr\"{o}m (DRN) solution is an electrostatic solution of the Einstein-Maxwell (EM) equations. In the axisymmetric description, the metric is defined by means of Weyl's line element \cite{Weyl}
\be ds^{2}=f^{-1}\left[e^{2\gamma}(d\rho^{2}+dz^{2})+\rho^{2}d\varphi^{2}\right]- f dt^{2},
\label{Weyl}\ee

\noi where $f$ and $\gamma$ are functions only of the cylindrical coordinates $(\rho,z)$. The corresponding Einstein-Maxwell field equations are given by
\be \bnabla \left(f^{-1} \bnabla A_{4}\right)=0, \qquad  f\Delta f=(\bnabla f)^{2} +
2f(\bnabla A_{4})^{2}, \label{fieldeq}\ee
\be 4\gamma_{\rho}=\rho f^{-2}(f^{2}_{\rho}-f^{2}_{z})-4\rho f^{-1}(A^{2}_{4\rho}-A^{2}_{4z}),\qquad
2\gamma_{z} = \rho f^{-2}f_{\rho}f_{z}-4\rho f^{-1}A_{4\rho}A_{4z}, \label{thegamma}\ee

\noi with $\Delta =\partial_{\rho\rho} + \rho^{-1} \partial_{\rho} + \partial_{z z} $ and $\bnabla = \univec_{\rho}  \partial_{\rho} + \univec_{z} \partial_{z}$, as the gradient and Laplace operators, respectively. $A_{4}$ is the electric potential, $\univec_{\rho}$ and $\univec_{z}$ are unit vectors. If we determine $A_{4}$ and $f$ from the coupled system of Eqs. (\ref{fieldeq}), then one integrates the system of Eqs. (\ref{thegamma}) and find later an explicit expression for the metric function $\gamma$. In order to accomplish this goal, first we combine Eqs. (\ref{fieldeq}) according to Ernst's formalism \cite{Ernst}
\be \left( {\cal{E}}+\Phi^{2}\right)\Delta{\cal{E}}=(\bnabla{\cal{E}}+ 2 \Phi \bnabla \Phi)\bnabla {\cal{E}}, \qquad  \left( {\cal{E}}+\Phi^{2}\right)\Delta \Phi=(\bnabla{\cal{E}}+
2 \Phi \bnabla \Phi)\bnabla \Phi, \label{ERNST}\ee

\noi where ${\cal{E}}$ and $\Phi$ are the Ernst potentials, which are defined by the following relations:
\be {\cal{E}}=f-A^{2}_{4}, \qquad \Phi=-A_{4}. \ee

The extreme DRN solution describes a binary system of two aligned Reissner-Nordstr\"{o}m sources interacting by means of gravitational and electric forces. In terms of canonical parameters $\{\alpha,\beta_{1},\beta_{2}\}$, a general extreme DRN solution is given by \cite{IMR}
\[{\cal{E}}=\frac{\Lambda+\Gamma}{\Lambda -\Gamma},\qquad \Phi=\frac{C}{\Lambda -\Gamma},\qquad
f= \frac{\Lambda^{2}-\Gamma^{2}+ C^{2}}{(\Lambda -\Gamma)^{2}},\qquad e^{2\gamma}= \frac{\Lambda^{2}-\Gamma^{2}+ C^{2}}{K^{2}_{0} r_{1}^{4} r_{2}^{4}}, \]
\[ \Lambda=\left|
\begin{array}{cccc}
\gamma_{11}r_{1} & \gamma_{12}r_{2} & -\gamma^{2}_{12}r^{2}_{2} & -\gamma^{2}_{11}r^{2}_{1} \\
\gamma_{21}r_{1} & \gamma_{22}r_{2} & -\gamma^{2}_{22}r^{2}_{2} & -\gamma^{2}_{21}r^{2}_{1} \\
M_{11} & M_{12} & W_{12} & W_{11} \\
M_{21} & M_{22} & W_{22} & W_{21} \\
\end{array}\right|, \quad \Gamma=\left|
\begin{array}{ccccc}
0 & 1 & 1 & P^{(2)}_{1} & P^{(1)}_{1} \\
1 & {} & {} & {} & {} \\
1 & {} & {} & \Lambda & {} \\
0 & {} & {} & {} & {} \\
0 & {} & {} & {} & {} \\
\end{array}
\right|, \quad
C=-\left|
\begin{array}{ccccc}
0 & f(\alpha_{1}) & f(\alpha_{2}) & F_{2}& F_{1}\\
1 & {} & {} & {} & {} \\
1 & {} & \Lambda & {} & {} \\
0 & {} & {} & {} & {} \\
0 & {} & {} & {} & {} \\
\end{array}
\right|,  \]
\[ K_{0}=\left|
\begin{array}{cccc}
\gamma_{11} & \gamma_{12} & -\gamma^{2}_{12} & -\gamma^{2}_{11} \\
\gamma_{21} & \gamma_{22} & -\gamma^{2}_{22} & -\gamma^{2}_{21} \\
M_{11} & M_{12} & \frac{\partial M_{12}}{\partial\alpha_{2}} & \frac{\partial M_{11}}{\partial\alpha_{1}} \\
M_{21} & M_{22} & \frac{\partial M_{22}}{\partial\alpha_{2}} & \frac{\partial M_{21}}{\partial\alpha_{1}} \\
\end{array}\right|, \quad P^{(k)}_{1}=\frac{z-\alpha_{k}}{r_{k}},\quad W_{ik}= r^{2}_{k} \frac{\partial}
{\partial \alpha_{k}} \left[\frac{M_{ik}}{r_{k}}\right],\quad
F_{k}= r^{2}_{k} \frac{\partial} {\partial \alpha_{k}} \left[\frac{f(\alpha_{k})}{r_{k}}\right],\]
\[M_{ik}=(e_{i}+ 2f_{i}f(\alpha_{k}))\gamma_{ik}, \quad f(\alpha_{k})=\sum^{2}_{l=1}f_{l}\gamma_{lk},
\quad \gamma_{ik}=\frac{1}{\alpha_{k}-\beta_{i}},\quad r_{k}=\sqrt{\rho^{2}+(z-\alpha_{k})^{2}},
\quad \alpha_{1}=-\alpha_{2}=\alpha, \quad i,k=1,2, \]
\be e_{l}=\frac{2(\beta^{2}_{l}-\alpha^{2})\left[(\beta^{2}_{l}-2\beta_{l}\beta_{m}+\alpha^{2})+
\epsilon_{l}\epsilon_{m}(\beta^{2}_{l}-\alpha^{2})\right]}{(\beta_{l}-\beta_{m})^{3}},
\quad f_{l}=\epsilon_{l} \frac{\beta^{2}_{l}-\alpha^{2}}{\beta_{l}-\beta_{m}},
\quad \epsilon_{l}=\pm 1, \quad l \neq m=1,2. \label{extremeDRN}\ee

The three canonical parameters $\{\alpha,\beta_{1},\beta_{2}\}$ in Eq. (\ref{extremeDRN}) can be asymptotically related with the physical Komar parameters, i.e., total mass $M$, electric charge $Q$, and a coordinate distance $R$ between the sources, by means of the following equations
\be e_{1} + e_{2}=-2M, \qquad  f_{1}+f_{2}= Q, \label{masscharge} \ee

\noi where the first Eq. (\ref{masscharge}) leads us to the following relation
\be \beta_{1}+\beta_{2}=-M.\label{betas} \ee

Using the above relation Eq. (\ref{betas}) and combining both Eqs. (\ref{masscharge}) we obtain the following algebraic equation
\be \Delta_{o}\left[(2P)^{2}-2(\Delta_{o}-2\alpha^{2})(2P)+M^{2}\Delta_{o}-4\alpha^{2}M^{2} + 4\alpha^{4} \right]=0, \qquad P:=\beta_{1} \beta_{2}, \label{theP}\ee

\noi where $\Delta_{o}=M^{2}-Q^{2}$ and $\alpha=R/2$. If $\Delta_{o}=0$, Eq. (\ref{theP}) reduces to the MP solution and if $\Delta_{o} \neq 0$ it becomes the complementary Bonnor's solution. A more suitable form of the above solution can be achieved in prolate spheroidal coordinates $(x,y)$, which are related to cylindrical coordinates $(\rho, z)$ via the formulae
\be x=\frac{r_{2}+ r_{1}}{2\alpha}, \quad y=\frac{r_{2}-r_{1}}{2\alpha}, \quad r_{1,2}=\sqrt{\rho^{2} +(z\pm\alpha)^{2}},\quad \Leftrightarrow \quad \rho=\alpha\sqrt{(x^{2}-1)(1-y^{2})}, \quad z=\alpha xy. \label{prolates}\ee

In terms of these new coordinates the line element reads
\be ds^{2}=\alpha^{2}f^{-1}\left[e^{2\gamma}(x^{2}-y^{2})\left(\frac{dx^{2}}{x^{2}-1}+\frac{dy^{2}}{y^{2}-1}\right)+
(x^{2}-1)(1-y^{2})d\varphi^{2}\right]- f dt^{2}\, .\label{Weylprolates}\ee

\noi Its explicit form in terms of physical parameters will depend on the sign of $\epsilon_{l}$ and of the functional form of $\beta_{l}$ in terms of Komar parameters \cite{Komar}.

\section{III. The Majumdar-Papapetrou two-body solution}
By setting in Eq. (\ref{masscharge}) $\epsilon_{1}=\epsilon_{2}=\epsilon=\pm1$, we recover the MP solution, since  $e_{l}=2\epsilon f_{l}$. Therefore $Q = -\epsilon M$ (or equivalently $\Delta_{o} = 0$). Hence the Ernst potentials and metric functions reduce to \cite{IMR}:
\[{\cal{E}}=\frac{E_{+}}{E_{-}}=1+2\epsilon \Phi,\qquad \Phi=\frac{F}{E_{-}},\qquad f=\frac{\alpha^{4}(x^{2}-y^{2})^{2}}{E_{-}^{2}}, \qquad  e^{2\gamma}=1,\]
\be E_{\pm}= \alpha^{2}(x^{2}-y^{2})\pm\alpha(\beta_{1}+\beta_{2})x\mp(\alpha^{2}+\beta_{1}\beta_{2})y, \qquad
F=\epsilon\left[\alpha(\beta_{1}+\beta_{2})x-(\alpha^{2}+\beta_{1}\beta_{2})y\right]. \label{MajumdarPapapetrou} \ee

As mentioned above, $\Delta_{o}=0$ reduces Eq. (\ref{theP}) to the MP solution. Nevertheless, one still needs to find the explicit form of the canonical parameters $\beta$ in terms of the physical Komar parameters. Manko \cite{Manko} presented explicit expressions of the canonical parameters as functions of the physical parameters, they read
\be\beta_{1,2}=- \frac{M_{1}+M_{2}}{2} \pm \frac{ \sqrt{(2\alpha - M_{1}+M_{2})^{2}+4M_{1}M_{2}}}{2}. \label{betasMP} \ee

Solving Eq. (\ref{sigmas}) in the extreme condition $\sigma_{j}=0$, the individual masses and charges result to be related by $Q_{1}=\pm M_{1}$ and $Q_{2}=\pm M_{2}$.

\subsection{Singular surfaces in the MP sector}
According with the positive mass theorem \cite{SchoenYau1,SchoenYau2}, the values of the individual masses must fulfill the condition $M>0$, for the total ADM mass \cite{ADM}. Nevertheless, since the positive mass theorem does not provide a feasible proof on the regularity of the solution, Eq. (\ref{MajumdarPapapetrou}), it could not be necessarily free of SS, even if the total ADM mass satisfies the condition $M=M_{1}+M_{2}>0$. Then, we should add analytic conditions that guarantee the regularity of the solution. In order to accomplish this goal, we use the physical representation of the denominator of the Ernst potentials, which is given by
\be F_{MP}=\left(x+\frac{M_{1}+M_{2}}{2\alpha}\right)^{2}-\left(y-\frac{M_{1}-M_{2}}{2\alpha}\right)^{2}
-\frac{M_{1}M_{2}}{\alpha^{2}}=0, \ee

\noi this is the equation of a hyperbola, whose asymptotes are given by
\be y=  \pm \left(x + \frac{M_{1}+M_{2}}{2\alpha}\right)+\frac{M_{1}-M_{2}}{2\alpha}. \ee

\begin{figure}[ht]
\begin{minipage}{0.49\linewidth}
\centering
\includegraphics[width=6cm,height=5cm]{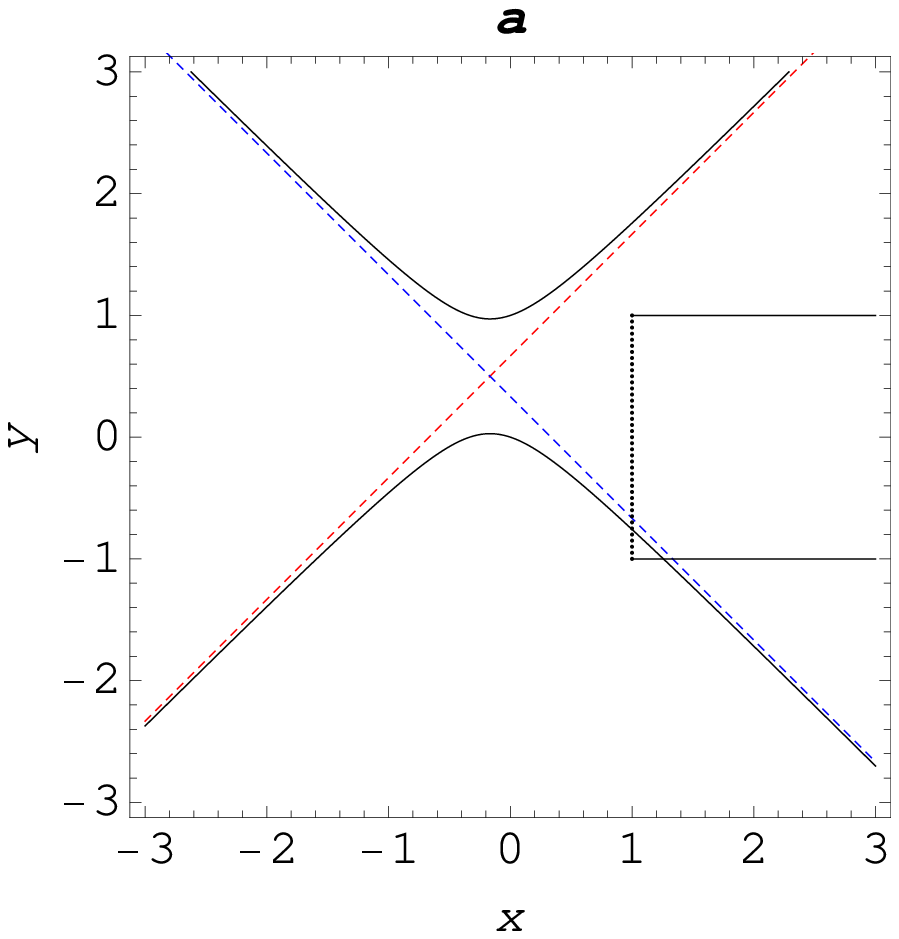}
\end{minipage}
\begin{minipage}{0.49\linewidth}
\centering
\includegraphics[width=6cm,height=5cm]{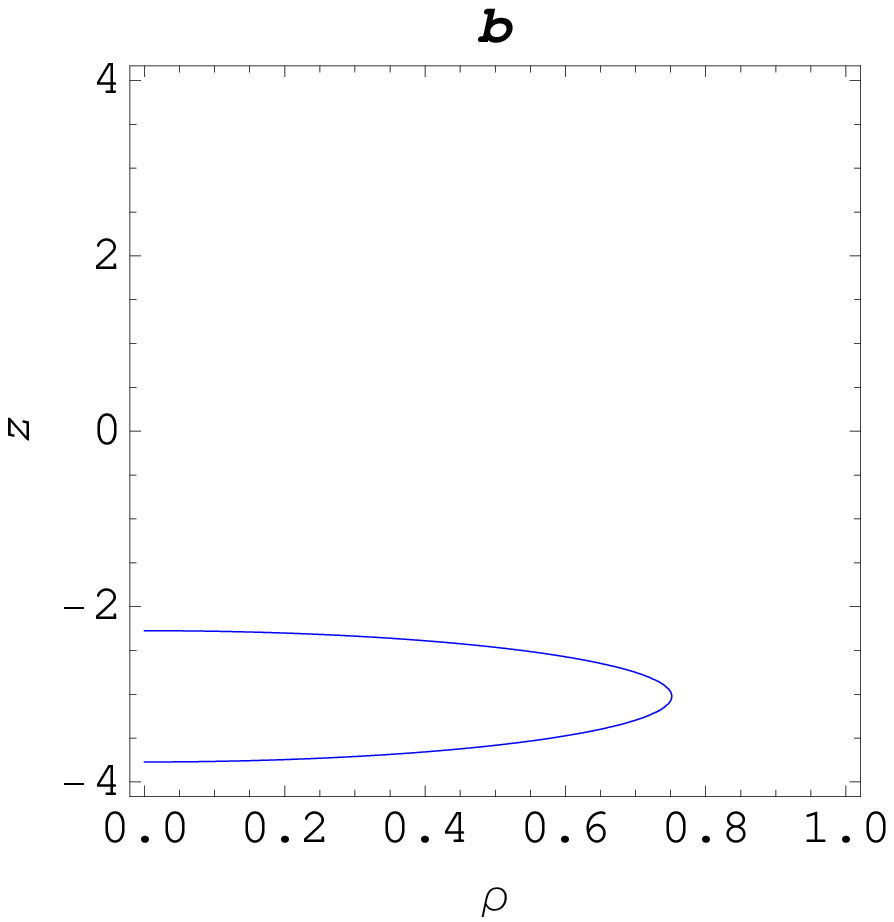}
\end{minipage}
\caption{(a) The crossing inside the region $x>1$, $|y|<1$, if one of the masses is negative (in this case $M_{2}<0$), for the values $\alpha=3$, $M_{1}=2$, and $M_{2}=-1$. (b) Appearance of the corresponding singular surface if $M_{2}<0$ in the MP sector.}
\label{SingularSurfacesMP}\end{figure}

Notice that the region defined by $x\geq1$ and $|y|<1$, shows the presence of SS off the axis [see Eq.(\ref{prolates})]. The conditions $x=1$ and $|y|< 1$ are sufficient to prove that at least one asymptote is crossing inside this region, therefore SS appear into the system under consideration. Without loss of generality, let us suppose that the straight line associated with the mass $M_{1}$ (the one with positive slope), is not crossing through this region, but the other one associated with the mass $M_{2}$ (the one with negative slope) does it (see Fig. \ref{SingularSurfacesMP}). Hence, we have that
\[  + \left( 1 + \frac{M_{1}+M_{2}}{2\alpha}\right)+\frac{M_{1}-M_{2}}{2\alpha}>1,\quad \Rightarrow \quad M_{1}>0, \]
\be -\left( 1 + \frac{M_{1}+M_{2}}{2\alpha}\right)+\frac{M_{1}-M_{2}}{2\alpha}>-1,\quad \Rightarrow \quad M_{2}<0.\ee

Additionally, if none of the straight lines crosses inside this region, there exist no SS off the axis. Therefore, one concludes that the formation of SS is due to the negative value of the individual masses.

\section{IV. Bonnor's solution }
A complementary description is achieved by setting in Eq. (\ref{masscharge}) $\epsilon_{1}=-\epsilon_{2}=\epsilon=\pm1$. Therefore, we find that $e_{l}\neq 2\epsilon f_{l}$ and $Q\neq-\epsilon M$ (or $\Delta_{o}\neq0$). In this case the Ernst potentials and metric functions are given by \cite{IMR}:
\[{\cal{E}}=\frac{E_{+}}{E_{-}},\qquad \Phi=\frac{\epsilon (\beta_{1}-\beta_{2})F}{E_{-}},\qquad f=\frac{\alpha^{4}(x^{2}-y^{2})^{2}}{E_{-}^{2}}, \qquad  e^{2\gamma}=\frac{N^{4}}{\alpha^{8}(\beta_{1}-\beta_{2})^{8}(x^{2}-y^{2})^{4}},\]
\[ E_{\pm}= \left[\alpha(\beta_{1}-\beta_{2})x-(\alpha^{2}-\beta_{1}\beta_{2})y \mp (\alpha^{2}- \beta_{1}^{2})\right]\left[\alpha(\beta_{1}-\beta_{2})x+(\alpha^{2}-\beta_{1}\beta_{2})y \pm (\alpha^{2}- \beta_{2}^{2})\right], \]
\[F=\alpha( 2\alpha^{2}- \beta_{1}^{2}-\beta_{2}^{2})x- (\beta_{1}+\beta_{2})(\alpha^{2}
-\beta_{1}\beta_{2})y,  \]
\be N=\alpha^{2}(\beta_{1}-\beta_{2})^{2}x^{2}- (\alpha^{2}-\beta_{1}\beta_{2})^{2}y^{2}+(\alpha^{2}-\beta_{1}^{2})(\alpha^{2}-\beta_{2}^{2}). \label{Bonnor} \ee

The explicit form of the canonical parameters $\beta$ in terms of the physical parameters can be obtained from
Eq. (\ref{theP}), excluding the MP case. A more general procedure is given in \cite{ILLM} in relation to identical Kerr-Newman sources. Cabrera-Munguia \emph{et al}. presented in Ref. \cite{IMR} their explicit form as a trivial consequence of the formulas presented by Manko \cite{Manko}, they read
\be\beta_{1,2}=-\frac{M_{1}+M_{2}}{2} \pm \sqrt{\frac{(2\alpha-M_{1}+M_{2})[4\alpha^{2}-(M_{1}+M_{2})^{2}]}{2\alpha+M_{1}-M_{2}}}\, .\label{betasB} \ee

Solving Eq. (\ref{sigmas}) for the extreme condition $\sigma_{j}=0$ and excluding the MP solution, we find that the individual masses and charges are related as follows
\be Q_{1}=\epsilon M_{1} \sqrt{\frac{(2\alpha+M_{2})^{2}-M_{1}^{2}}{(2\alpha-M_{2})^{2}-M_{1}^{2}}}, \qquad
Q_{2}= -\epsilon M_{2} \sqrt{\frac{(2\alpha+M_{1})^{2}-M_{2}^{2}}{(2\alpha-M_{1})^{2}-M_{2}^{2}}}, \qquad \epsilon=\pm 1. \ee

\subsection{Singular surfaces in Bonnor's sector}
As we did in the MP sector, we can prove also in this sector the conditions ensuring that the solution Eq.(\ref{Bonnor}) is free of SS and therefore regular off the axis. The polynomial contained in the denominator of the Ernst potentials has the following form:
\be F_{B}=D^{2}\left(x+\frac{M_{1}+M_{2}}{2\alpha}\right)^{2}-\left(y-\frac{M_{1}-M_{2}}{2\alpha}\right)^{2}=0,\qquad D:= \sqrt{\frac{4\alpha^{2}-(M_{1}-M_{2})^{2}}{4\alpha^{2}-(M_{1}+M_{2})^{2}}}, \ee

\noi which represents the geometric locus of two straight lines intersecting at an angle of
\be \theta=2\arctan D, \ee

\noi the straight lines are given by
\be y= \pm D \left( x + \frac{M_{1}+M_{2}}{2\alpha}\right)+ \frac{M_{1}-M_{2}}{2\alpha}. \ee

\begin{figure}[ht]
\begin{minipage}{0.49\linewidth}
\centering
\includegraphics[width=6cm,height=5cm]{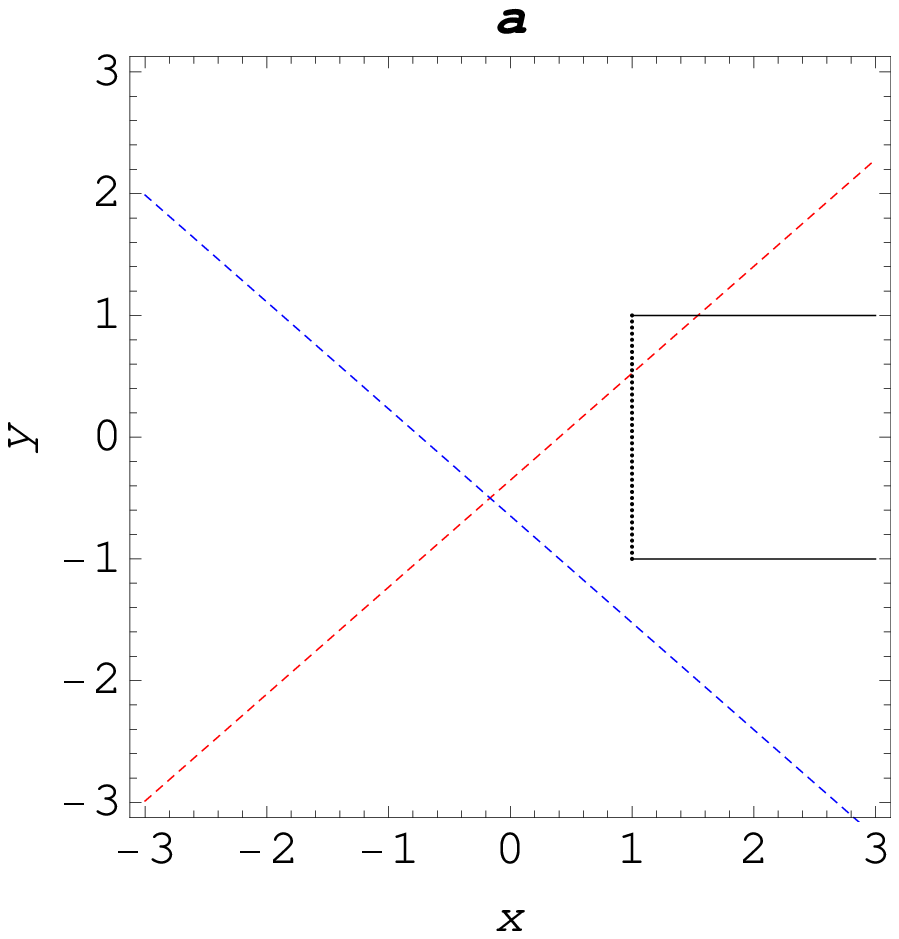}
\end{minipage}
\begin{minipage}{0.49\linewidth}
\centering
\includegraphics[width=6cm,height=5cm]{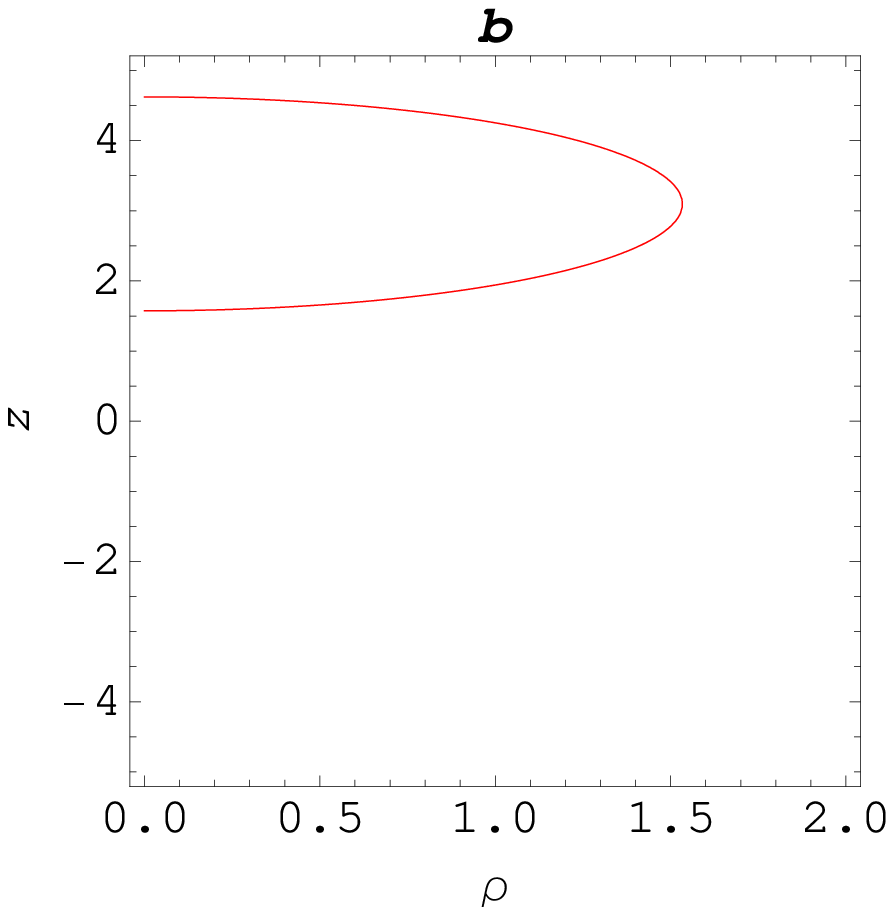}
\end{minipage}
\caption{(a) The crossing inside the region $x>1$, $|y|<1$, if one of the masses is negative (in this case $M_{1}<0$), for the values $\alpha=3$, $M_{1}=-1$, and $M_{2}=2$. (b) Appearance of the corresponding singular surface if $M_{1}<0$ in Bonnor's sector.}
\label{SingularSurfacesBonnor}\end{figure}

Once again, we notice that the conditions $x=1$ and $|y|< 1$ are sufficient to prove that at least one of the straight lines is crossing inside this region, hence forming SS off the axis (see Fig. \ref{SingularSurfacesBonnor}). We have that
\[ + D\left( 1 + \frac{M_{1}+M_{2}}{2\alpha}\right)+\frac{M_{1}-M_{2}}{2\alpha}>1,\quad \Rightarrow \quad \sqrt{\frac{(2\alpha+M_{1})^{2}-M_{2}^{2}}{(2\alpha-M_{1})^{2}-M_{2}^{2}}}>1, \quad \Rightarrow \quad M_{1}>0, \]
\be -D\left( 1 + \frac{M_{1}+M_{2}}{2\alpha}\right)+\frac{M_{1}-M_{2}}{2\alpha}>-1,\quad \Rightarrow \quad \sqrt{\frac{(2\alpha+M_{2})^{2}-M_{1}^{2}}{(2\alpha-M_{2})^{2}-M_{1}^{2}}}<1, \quad \Rightarrow \quad M_{2}<0.\ee

Moreover, there exist no SS off the axis if the straight lines do not cross inside this region.

\section{V. Conclusions}
In this work we present the first analytical proof of the conditions the extreme double Reissner-Nordstr\"{o}m solution should satisfy in order to avoid or develop singular surfaces off the axis, i.e., naked singularities. It is worthwhile to mention that the positive mass theorem establishes that a regular solution contains a total positive ADM mass. Nevertheless, the positiveness of the total mass cannot be the unique condition to prove that the solution is free of singularities anywhere. One must look at the denominator of the Ernst potentials in order to derive the analytical conditions which avoid the formation of SS into the solution. We expect to accomplish a deeper analysis about this subject in these binary systems, and other ones including the rotation parameter. The study of naked singularities is certainly quite intriguing and deserves further investigations.

\section{Acknowledgments.}
This work was supported by CONACyT Grant No. 166041F3 and by CONACyT fellowship with CVU No. 173252.

\end{document}